\documentclass[prd,a4paper,twocolumn,twoside,showpacs,amssymb]{revtex4}

\usepackage{latexsym}
\usepackage{dcolumn}
\usepackage{graphicx}
\usepackage{subfigure}
\usepackage{longtable}
\usepackage{amsmath}
\usepackage{supertabular}
\usepackage{ccaption}
\usepackage{natbib}
\usepackage[latin1]{inputenc}
\usepackage{amssymb}
\usepackage{graphics}
\usepackage{amssymb}
\usepackage{amsfonts}
\usepackage{bm}
 
\begin{document}

\pretolerance 10000

\newcommand*{\be}{\begin{equation}}
\newcommand*{\ee}{\end{equation}}
\newcommand{\beq}{\begin{equation}}
\newcommand{\eeq}{\end{equation}}
\newcommand{\ove}{\overline}
\newcommand{\half}{\frac 1 2 }
 
\title{Cosmological redshift and nonlinear electrodynamics propagation of photons 
from distant sources}
 
\author{Herman J. Mosquera Cuesta$^{1,2}$, Jos\'e M. Salim$^1$ and M. Novello$^1$}
 
\affiliation{$^1$\mbox{Instituto de Cosmologia, Relatividade e Astrof\'{\i}sica 
(ICRA-BR), Centro Brasileiro de Pesquisas F\'{\i}sicas (CBPF)} \\ \mbox{Rua Dr.  
Xavier Sigaud 150, Urca 22290-180, RJ, Brasil :::  hermanjc@cbpf.br}  }
 
\date{\today}

\begin{abstract}
By-now photons are the unique universal messengers. Cosmological sources like far-away 
galaxies or quasars are well-known light-emitters. Here we demonstrate that the nonlinear 
electrodynamics  (NLED)  description of photon propagation  through the weak background 
intergalactic magnetic fields modifies in a fundamental way the cosmological redshift that 
a direct computation within a specific cosmological model can abscribe to a distant source. 
Independently of the class of NLED Lagrangian, the effective redshift turns out to be  
$1 + \tilde{z} = (1 + z)~\Delta$, where $\Delta \equiv (1 + \Phi_e)/(1 + \Phi_o)$, 
with $\Phi \equiv {8}/{3} ({L_{FF}}/{L_F}) B^2$, being $L_F = {dL}/{dF}$, $L_{FF} = {d^2L
}/{dF^2}$, the field $F\equiv F_{\alpha \beta} F^{\alpha \beta}$, and $B$ the magnetic 
field strength. Thus the effective redshift is always much higher then the standard redshift, 
but recovers such limit when the NLED correction $\Delta(\Phi_e, \Phi_o) \longrightarrow 1$. 
This result may provide a physical foundation for the current observation-inspired interpretation 
that the universe undergoes an accelerate expansion. However, under the situation analyzed here,  
for any NLED the actual (spatial) position of the light-emitting far-away source remains untouched. 

\end{abstract}



\maketitle



{\sl Introduction.---} The study of the expansion history of the universe gained a novel 
dimension after the discovery of what appears to be a dimming in the luminosity emitted by 
supernovae type Ia (SNIa) \cite{riess98}, which are thought of as standard candles. These 
observations have been interpreted in the context of the standard  cosmological model as 
evidence of a late-time transition from decelerate-to-acelerate expansion, according 
to most current viewpoints. The conclusion is attained after combining both the redshift
and luminosity-distance of observed SNIa events in their Hubble diagram (HD). If one excludes 
any potential systematics (as for instance, there exists the possibility that we are being 
unable to detect, due to dust effects, much more reddened SNIa taking place at much higher 
redshifts ($z$), than close-by bluish explosions because simply these late ones are much 
brighter, see Ref.\cite{della-valle}), one can verify that their representative points in the 
HD appear a bit over the upper bound curve predicted by the standard Friedmann cosmology, and 
pile-up around $z \sim 0.5-1$, which is referred to as the transition era.

The redshift is the fundamental piece in achieving this conclusion. It is determined in almost 
all the cases by analysing the absorption lines from SNIa host galaxies. However, the unavoidable 
nonlinear interaction of light \cite{burke97} from these distant sources with the intergalactic 
background magnetic fields \footnote{In Ref.\cite{burke97} has been proved that electrodynamics 
in a vacuum pervaded by B-fields is subject to nonlinear effects. } may crucially modify the 
putative value of the redshift to be abscribed to a given source from the observed lines. On the 
other hand, exception done for the case in which the electric permittivity and magnetic permeability 
are functions of the fields, i.e., $\epsilon(\vec{E}, \vec{B})$, $\mu(\vec{E}, \vec{B})$, Maxwell 
electrodynamics is unable to describe the nonlinear 
behavior of light propagation. If one follows this line of reasoning, one realizes that one way to 
guide ourselves to a proper investigation of the nonlinear interaction of photons \cite{burke97} from 
distant galaxies and quasars with intergalactic background fields is to keep in mind that those 
background magnetic fields are extremely weak! (observations rule out any electric fields, i. e., 
$<E>=0$!). Hence, if a given Lagrangian will indeed describe such photon nonlinear dynamics, it will 
have to depend on the invariant $F \equiv F_{\mu \nu} F^{\mu \nu}$ field in a nontrivial fashion
Interestingly, a hint to the need for a nonlinear 
electrodynamics (NLED) Lagrangian able to account for such dynamics came to us from the study 
of a correlated phenomenon: claims on a potential variation of the fine structure constant 
$\alpha$ (see \cite{jean-paul-herman07} and references therein). Indeed, Murphy et al.
\cite{murphy01} (Section 2.6) based on Maxwell 
electromagnetic theory considered large magnetic fields as a potential cause of systematic 
errors in their measurements of $\Delta \alpha/\alpha$, and conclude that the intra-cluster 
magnetic field strengths are nine orders of magnitude below the strength required to cause 
substantial effects. Therefore, it appears legitimate to address the question of any putative 
modification of the standard cosmological redshift from far away astrophysical sources within 
the framework of NLED. One way to achieve this goal is to use the Lagrangian for NLED recently 
introduced in Ref.\cite{novello04} (see also Ref.\cite{nos2006}), 
whose original focus was to bring in an alternative to dark energy to explain the universe late-time 
accelerate expansion. In passing, it turns to be easy to check that Murphy et al.'s conclusion can 
be reversed by considering the NLED Lagrangian density of Ref.\cite{novello04}.



{\sl Nonlinear electromagnetism in Cosmology.---} In Ref.\cite{novello04} several general 
properties of nonlinear electrodynamics in cosmology were reviewed by assuming that the 
action for the electromagnetic field is that of Maxwell with an extra term, namely
\footnote{Notice that this Lagrangian is gauge invariant, and that hence charge 
conservation is guaranteed in this theory.}
\beq
S = \int \sqrt{-g} \left( - \frac F 4 + \frac \gamma F \right) d^4x \; ,
\label{action}
\eeq
where $F\equiv F_{\mu\nu}F^{\mu\nu}$. Physical motivations for bringing in this theory
have been provided in Ref.\cite{novello04}.
At first, one notices that for high values of the field $F$, the dynamics resembles 
Maxwell's one except for small corrections associate to the parameter $\gamma$, while 
at low strengths of $F$ it is the $1/F$ term that dominates \cite{gamma}. (Clearly, this 
term should dramatically affect the photon-$\vec{B}$ field interaction in intergalactic 
space). The consistency of this theory with observations was shown in Refs.\cite{novello04,
nos2006} using the cosmic microwave radiation bound and the anomaly in the dynamics of 
Pioneer spacecraft, respectively. Both provide small enough values for the parameter 
$\gamma$. 

Therefore, the electromagnetic (EM) field described by Eq.(\ref{action}) can be taken as 
source in Einstein equations, to obtain a toy model for the evolution of the 
universe which displays accelerate expansion caused when the nonlinear EM term takes over 
the term describing other matter fields. This NLED theory yields ordinary radiation plus 
a dark energy component with $w < -1$ (phantom-like dynamics). Introducing the notation
\footnote{Due to the isotropy 
of the spatial sections of the Friedman-Robertson-Walker (FRW) model, an average procedure 
is needed if electromagnetic fields are to act as a source of gravity \cite{tolman-ehrenfest}. 
Thus a volumetric spatial average of a quantity $X$ at the time $t$ by $\langle X \rangle_{|_V} 
\equiv \lim_{V\rightarrow V_0} \frac 1 V \int X \sqrt{-g}\;d^3x$, where $V = \int \sqrt{-g} 
\;d^3x$ , and $V_0$ is a sufficiently large time-dependent three-volume. (Here the metric sign 
convention $(+---)$ applies).}, the EM field can act as a source for the FRW model 
if $\langle E_i \rangle_{|_V} =0, \; \langle B_i \rangle_{|_V} =0,\; \langle {E_i B_j} \rangle_
{|_V} = 0$, $ \langle {E_iE_j} \rangle_{|_V} = - \frac 1 3 E^2 g_{ij}$, and $\; \langle {B_iB_j} 
\rangle_{|_V} = -\frac 1 3 B^2 g_{ij}$.\footnote{Let us remark that since we are assuming that 
$\langle {B}_i \rangle_{|_V} = 0$, the background magnetic fields induce no directional effects 
in the sky, in accordance with the symmetries of the standard cosmological model.} When these 
conditions are fulfilled, a general nonlinear Lagrangian $L(F)$ yields the energy-momentum tensor
($L_F = {dL}/{dF}, \;\; L_{FF} = {d^2L}/{dF^2}$)\footnote{Under the same 
assumptions, the EM field associate to Maxwell Lagrangian generates the stress-energy tensor 
defined by Eq.(\ref{tmunu}) but now $ \rho = 3 p = \half(E^2 + B^2)$.}  
\begin{eqnarray}
& \langle {T}_{\mu\nu} \rangle_{|_V}  =  (\rho + p) v_\mu v_\nu - p\; g_{\mu\nu} , & \\ 
& \rho  =  -L - 4E^2 L_F , \;\;\;\;\; p =  L + \frac 4 3 (E^2-2B^2) L_F , & \nonumber
\label{tmunu}
\end{eqnarray}
Hence, when there is only a magnetic field, the fluid can be thought of as 
composed of ordinary radiation with $p_{1}= \frac 1 3\; \rho_{1}$ and of another fluid with 
EOS $p_{2} = -\frac 7 3 \;\rho_{2}$. It is precisely this component with negative pressure 
that may drive accelerate expansion.

After presenting that theory in Ref.\cite{novello04}, we realized that there exists another {\sl 
per se} equally fundamental implication of this Lagrangian. It also modifies in a significant 
fashion the actual redshift that one may adscribe, within a given cosmology, to a distant galaxy. 




{\sl Photon dynamics in NLED: effective geometry.---}        Next we investigate the effects of 
nonlinearities in the evolution of EM waves in the vacuum permeated by background $\vec{B}$-fields. 
An EM wave is described onwards as the surface of discontinuity of the EM field. Extremizing the 
Lagrangian $L(F)$, with $F(A_\mu)$, with respect to the potentials $A_{\mu}$ yields the following 
field equation \cite{plebanski}
\be 
\nabla_{\nu} (L_{F} F^{\mu\nu} ) = 0\label{eq60} , 
\ee
where $\nabla_\nu $ defines the covariant derivative. Besides this, 
we have the EM field cyclic identity
\be
\nabla_{\nu} F^{*\mu\nu} = 0 \hskip 0.3 truecm \Leftrightarrow 
\hskip 0.3 truecm F_{\mu\nu|\alpha} + F_{\alpha\mu|\nu} + 
F_{\nu\alpha|\mu} = 0\; . 
\label{eq62}
\ee
Taking the discontinuities of the field Eq.(\ref{eq60}) one gets \footnote{Following 
Hadamard \cite{HAD}, the surface of discontinuity of the EM field is denoted by $\Sigma$. 
The field is continuous when crossing $\Sigma$, while its first derivative presents a finite 
discontinuity. These properties are specified as follows: $\left[F_{\mu \nu} \right]_{\Sigma} 
= 0\; ,$ \hskip 0.3 truecm $\left[F_{\mu\nu|\lambda}\right]_{\Sigma} = f_{\mu\nu} k_\lambda\; 
\protect \label{eq14} \;$, where the symbol $\left[F_{\mu \nu}\right]_{\Sigma} = \lim_{\delta 
\to 0^+} \left(J|_{\Sigma + \delta}-J|_{\Sigma - \delta}\right)$ represents the discontinuity 
of the arbitrary function $J$ through the surface $\Sigma$. The tensor $f_{\mu\nu}$ is called 
the discontinuity of the field,  $k_{\lambda} = \partial_{\lambda} \Sigma $ is the propagation 
vector, and the symbols "$_|$" and "$_{||}$" stand for partial and covariant derivatives.} 
\be 
L_{F} f^{\; \; \mu}_{\lambda} k^\lambda + 2L_{FF}F^{\alpha\beta} 
f_{\alpha\beta} F^{\mu\lambda} k_{\lambda} = 0 \; ,
\label{j1} 
\ee
which together with the discontinuity of the Bianchi identity\footnote{A cyclic identity for 
the first derivative of the Riemann tensor, defined as: $R^\alpha_{\; \; \beta \mu \nu; \sigma} 
+ R^\alpha_{\; \; \beta \nu \sigma ; \mu } + R^\alpha_{\; \; \beta  \sigma \mu ; \nu} = 0$ } 
yields
\be
f_{\alpha\beta}k_{\gamma} + f_{\gamma\alpha}k_{\beta} + 
f_{\beta \gamma} k_{\alpha} = 0\; .
\ee
A scalar relation can be obtained if we contract this equation
with $ k^{\gamma}F^{\alpha\beta} \label{eq25}$, which yields
\be
(F^{\alpha\beta}f_{\alpha\beta} g^{\mu\nu} + 2F^{\mu\lambda} 
f_{\lambda}^{\; \; \nu})k_{\mu} k_{\nu}=0 \; .
\label{j2}
\ee
It is straightforward to see that here we find two distinct solutions: a) 
when $F^{\alpha\beta} f_{\alpha\beta}=0$, case in which such mode propagates 
along standard null geodesics, and b) when $F^{\alpha\beta} f_{\alpha\beta}
=\chi$. In this last case, we obtain from equations (\ref{j1}) and (\ref{j2}) 
the propagation equation for the field discontinuities being given by 
\cite{novello2000}
\be
\underbrace{ \left(g^{\mu\nu} - 4\frac{L_{FF}}{L_{F}} F^{\mu\alpha} 
F_{\alpha}^{\; \; \nu}\right) }_{\rm effective \; metric} k_{\mu}k_{\nu} = 0 \; .
\label{63}
\ee
This equation proves that photons propagate following a geodesic that is
not that one on the background space-time $g^{\mu\nu}$, but rather they 
follow the {\sl effective metric } given by Eq.(\ref{63}), which depends 
on the background $F^{\mu\alpha}$, i. e., on the $\vec{B}$-field.         

If one now takes the derivative of this expression, we can easily
obtain \cite{herman-salim06,herman-salim04,herman-salim04A}
\be
k^{\nu} \nabla_{\nu} k_{\alpha}  = 4 \left(\frac{L_{FF}}{L_{F}}F^{\mu\beta}
F_{\beta}^{\; \; \nu} ~k_{\mu} k_{\nu}\right)_{|\alpha}.
\label{kuknu}
\ee
This expression shows that the nonlinear Lagrangian introduces a term acting as a force 
that accelerates (positively or negatively) the photon along its path. 

It is therefore essential to investigate what  are the effects of this peculiar photon 
dynamics. The occurrence of this phenomenon over cosmological distance scales may have 
a nonnegligible effect on the physical properties that one can abscribe to a given source 
from astronomical observables. One example of this is the cosmological redshift, i.e., 
the actual shifting in the position of the absorption lines from far away galaxies. 
\footnote{The same theory directly leads to a variation of the fine structure constant 
$\alpha$\cite{jean-paul-herman07}}.
Indeed, it is proved next that as the photon travels very long distances from cosmic sources 
until be detected on Earth, its interaction with the background electromagnetic fields pervading 
intergalactic space decidedly changes the putative value of the redshift to be abscribed to the 
emitting source in case the NLED effects were not considered. In other words, the redshift computed 
in a standard fashion within a particular cosmology gets effectively affected.



Above we obtained the effective contravariant metric (see Ref.\cite{novello2000})
\be
g^{\mu\nu}_{\rm eff} \equiv L_F g^{\mu\nu} - 4 L_{FF} F^{\mu\alpha} F_{\alpha}^{\; \;\nu} \;. 
\label{63A}
\ee
Since we assume that the fields detected by a comoving observer is
\be
< F_{\mu \alpha} F^{\alpha}_{\; \; \;\nu} > = - \frac{1}{3} B^2 h_{\mu \nu} 
\label{average-field}
\ee
then, one can rewrite Eq.(\ref{63A}) explicitly in terms of the average magnetic field on the 
background as 
\begin{eqnarray}
g^{\mu\nu}_{\rm eff}  =  L_F g^{\mu\nu} + \frac{8}{3} L_{FF} B^2 h^{\mu \nu} \; .
\label{64}
\end{eqnarray}
On this basis, the inverse effective metric should be 
\begin{eqnarray}
g_{\mu \nu}^{\rm eff} = \frac{1}{L_F} g_{\mu \nu} - \frac{8}{3} \frac{ L_{FF} B^2 }{ L_F 
\left(L_F + \frac{8}{3} L_{FF} B^2 \right)} h_{\mu \nu}  \; .
\label{covar-eff-metric}
\end{eqnarray}
That theory leads straightforwardly to prove that the cosmological redshift of a photon 
traveling from a distant source to Earth is also modified.

{\sl Nonlinear cosmological redshift.---} With this metric one can 
then compute the new cosmological redshift since the line element provided by this NLED 
now reads
\begin{widetext} 
\begin{eqnarray}
ds^2 = \frac{1}{L_F} dt^2 - \left( 1 - \frac{8 }{3} \frac{L_{FF } B^2}{\left[L_F + 
\frac{8}{3} B^2 L_{FF} \right]}\right) \frac{a^2(t)}{L_F} \gamma_{ij} d{\rm x}^i 
d{\rm x}^j = \frac{1}{L_F} dt^2 - \frac{1}{L_F} \left( \frac{3 L_F}{3 L_F + 8 L_{FF} 
B^2} \right) a^2(t) dl^2 = 0 \; .
\label{photon-distance}
\end{eqnarray}
\end{widetext}
It follows then that the expression for the cosmological redshift turns out to be
\beq
1 + \tilde{z} \equiv \frac{c \delta t_0}{c \delta t_e} =  \frac{a(t_0)}{a{(t_e)}}~\Delta  
= (1 + z)~\Delta \; ,
\label{actual-redshift}
\eeq
where $\Delta \equiv [{\left(1 + \Phi \right)^{1/2} \big|_{t_e} }]/[{\left( 
1 + \Phi \right)^{1/2} \big|_{t_0} }]$, and $\Phi \equiv {8}/{3} ({L_{FF}}/{L_F}) B^2$.
The specific modification of the redshift depends on the particular problem we focus on, in 
particular, a similar effect was already analyzed in the presence of very strong magnetic 
fields in pulsars \cite{herman-salim06,herman-salim04,herman-salim04A}.

As an example, in the case of cosmology, as pointed out above, a model to explain the recently 
discovered 
late acceleration of the universe using a NLED described by the Lagrangian  $L(F) = - \frac{1}
{4} F  + \frac{\gamma}{F}$, where $F \equiv F_{\alpha \beta} F^{\alpha \beta}$ and $\gamma = - 
\nu^2$, was proposed in Ref.\cite{novello04}. Thus, by using Eqs.(\ref{covar-eff-metric}, 
\ref{actual-redshift}), one can compute the actual redshift of a given cosmic source. 

Therefore, the cosmological redshift turns out to be
\beq
1 + z = \frac{c \delta t_0}{c \delta t_e} =  \frac{a(t_0)}{a{(t_e)}} 
\frac{ {\left(\frac{B^4 + \frac{5}{3} \nu^2}{B^4 - \nu^2} \right)^{1/2} \Big|_{t_e} } }
{ {\left( \frac{B^4 + \frac{5}{3} \nu^2}{B^4 - \nu^2} \right)^{1/2} \Big|_{t_0} } } \; .
\label{actual-redshift-2}
\eeq
With this equation one can plot the effective cosmological redshift according to NLED for 
a specific field strength $B$. That result is presented in Fig.1, which illustrates through 
the Hubble diagram, a noticeable effect on both supernovae (SNe) and afterglows of gamma-ray 
bursts (GRBs) redshifts. Therefore, the determination from direct observations of the redshift 
of a given distant quasar or galaxy leads to a mistaken interpretation on the actual distance 
to those sources if such NLED effect is not properly accounted for. Such a task could be 
performed by estimating the host-galaxy $B$-field. 


\begin{figure*}[hbt]
\centering
\includegraphics[scale=0.21]{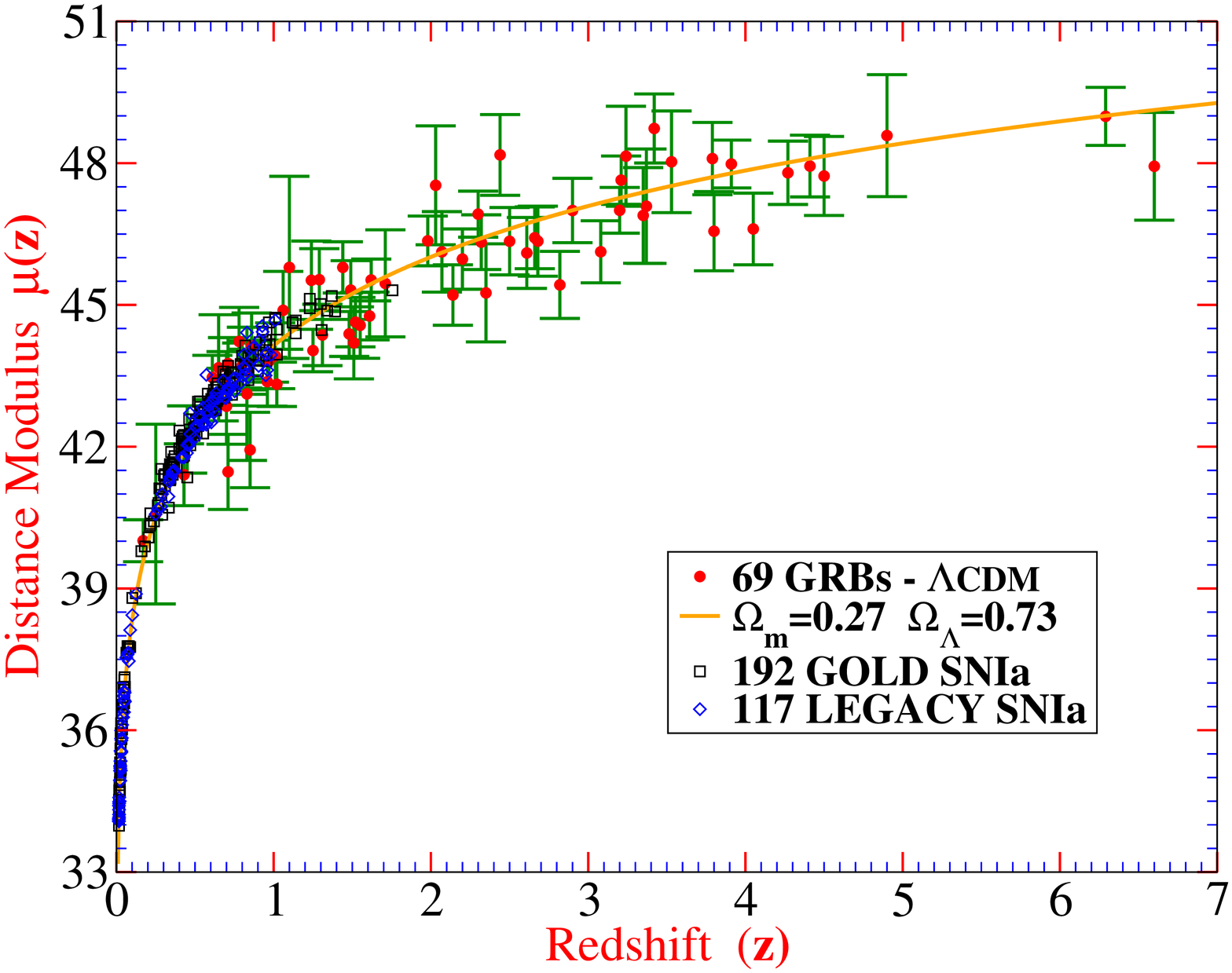}
\hskip 0.25truecm
\includegraphics[scale=0.24]{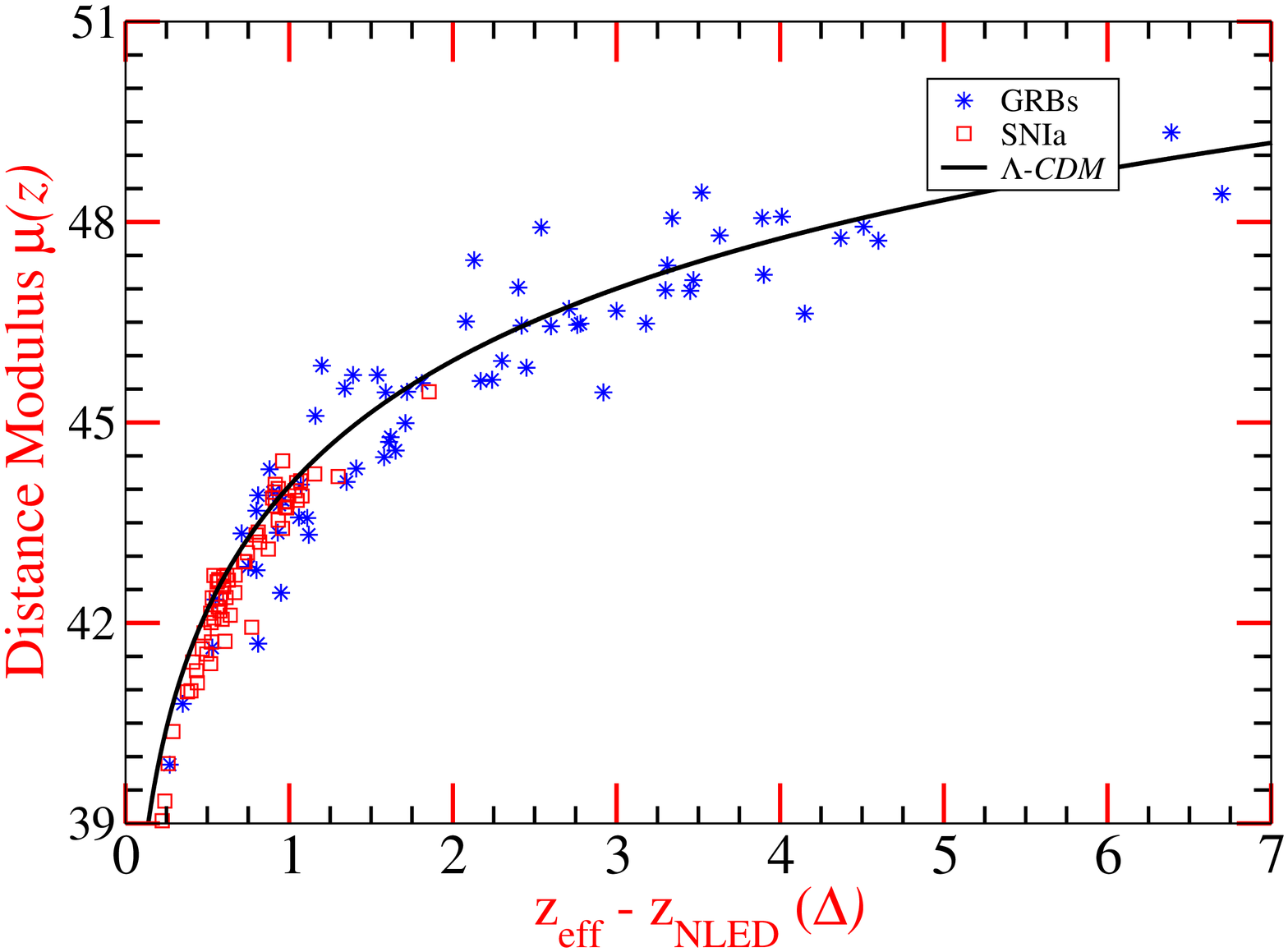}
\hskip 0.25truecm
\includegraphics[scale=0.24]{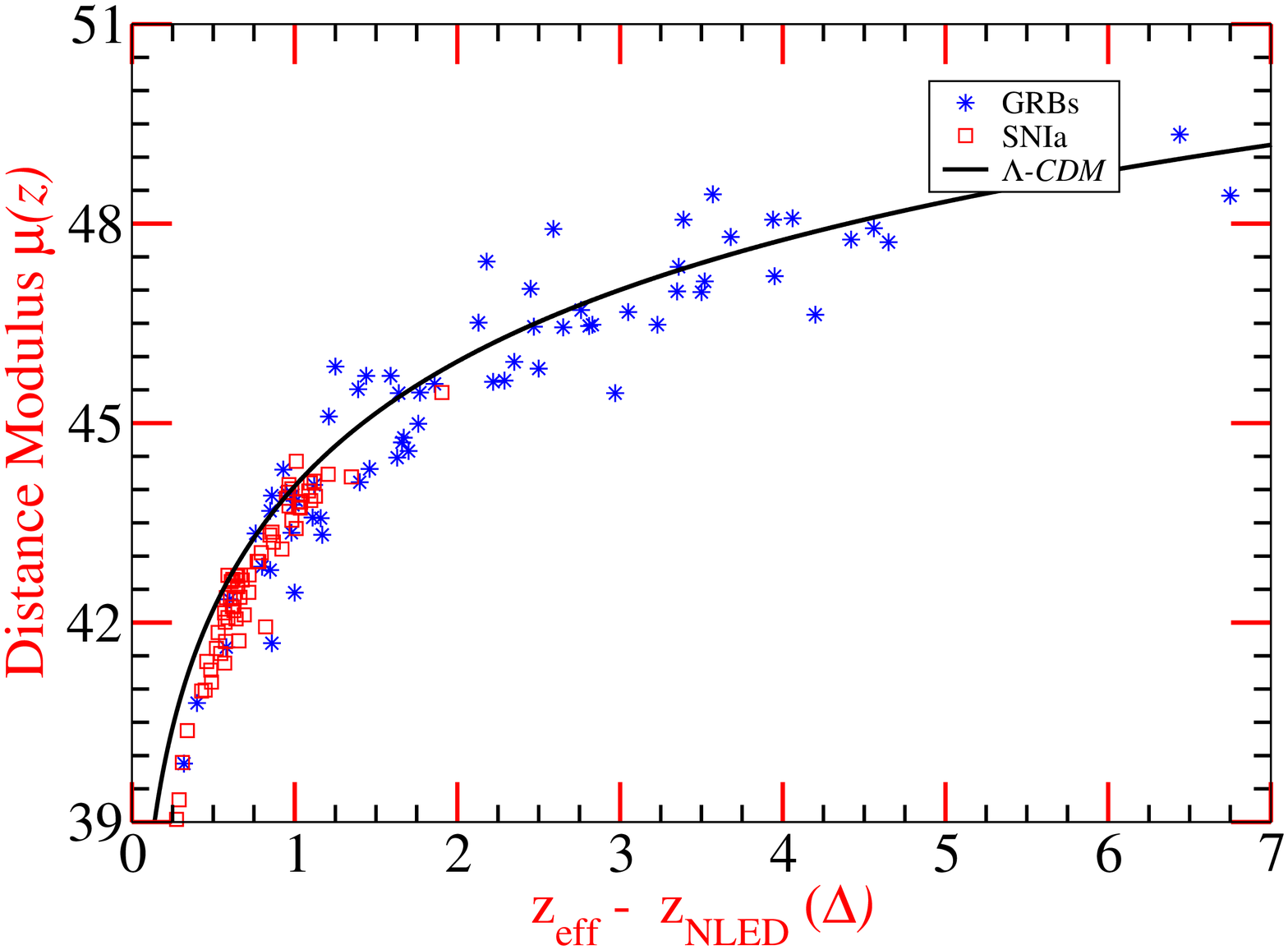}
\caption{Left figure: Curve (blue) delineating the NLED correction (CF$_{\rm NLED}$) to the 
standard cosmological redshift $(1+z)$ vs. the intergalactic $B$-field strength normalized 
as $B^4/\gamma$. 
Right figure:  Hubble diagram of the 69 GRBs sample (filled circles) of Ref.\cite{schaefer07} 
calibrated for $\Lambda$-CDM, 192 GOLD SNIa (squares) and 117 LEGACY SNIa (diamonds) samples 
of Refs.\cite{riess98,hungaros}, as current observations indicate. One may conjecture that most, 
if not all, of the events presenting a much higher luminosity in this plot could have taken place 
in sources where the local $B$-field is near the critical one shown at the left. The next two 
graphs illustrate SNIa, GRBS global shifts of 0.1, 0.15 in $z$. They are intended here solely 
to illustrate the overall effect. The attentive reader should bear in mind that the NLED 
correction should indeed be applied to each individual SNIa or GRB event, once knowing (through 
Zeeman splitting or other techniques) the local host-galaxy $B$-field. This analysis will be 
presented elsewhere.} 
\label{figure3} 
\end{figure*}

{\sl Discussion and conclusion.---} 
Our main result is presented in Eq.(\ref{actual-redshift}). Irrespective of the structure of the 
Lagrangian, it is valid for any generic Lagrangian $L(F)$ describing a NLED theory, provided the 
field averaging procedure indicated earlier in Eq.(\ref{average-field}) is hold. Therefore, in 
order to properly address the issue on the cosmological redshfit of a distant astrophysical object, 
this NLED effect should be first taken into account. In the specific case of the model introduced 
in Ref.\cite{novello04} to provide an alternative to dark energy to explain the current universe 
acceleration phase, Eq.(\ref{actual-redshift}) becomes Eq.(\ref{actual-redshift-2}) which shows 
the dependence of the NLED effect on the relation between the host source $B$-field strength and 
the theory constant $\gamma$, as shown in Fig.-\ref{figure3}.

As a simple exercise, one can admit for the time being that the $|_{t=0}$ (on Earth) $B$-field 
strength is much higher than the local one at the emitting source $|_{t=e}$. Since $\gamma$ is 
a universal constant of fixed value, then, in this limit one notices that the second term in the 
denominator of Eq.(\ref{actual-redshift-2}) is on the order of 1. \footnote{Ref.\cite{nos2006} 
has shown that $|\gamma|^{1/4}$ $\left(\frac{1}{c}\right) = B_{\rm crit}$!} By comparing to the 
standard calculation within a given model, {\sl viz.}, in Friedmann cosmology, one notices that 
for values of the ratio $B^4/\gamma$ in the interval $(1, +\infty)$, one obtains much higher 
redshifts, having the standard cosmological redshift recovered in the limit $B^4 >> \gamma$. 
These findings may provide a physical support to the observation-inspired interpretation that 
the universe is currently undergoing an accelerate expansion. Nonetheless, in any NLED theory, 
the actual position of the light-emitting far-away source remains unaffected! {Thus, a 
nonnegligible affect on the claimed late-time acceleration phase appears to occur.} 

On the other hand, notice that for $B^4 \longrightarrow |\gamma|$, the effective redshift $(1
+ \tilde{z}) = a{(t_0)}/a{(t_e)}~\Delta$ appears to diverge. That is, the source 
may appear infinitely redshifted! This seems to be a realistic possibility. Indeed it is not
\footnote{We caution on this apparent possibility to happen, because the procedure that was 
used to estimate the value of 
$\gamma$, and its associate $B_{\rm crit}$ critical field, in Ref.\cite{nos2006}, already 
includes uncertainties in the average intergalactic $B$-field, which are very large, yet. Of 
course, a more consistent fashion to obtain $\gamma$ would be through a dedicate laboratory 
experiment, as already stated in Ref.\cite{nos2006}. In that case, such very self-consistent 
method should produce a much smaller value of $\gamma$ than the one computed and described in 
footnote-3 of Ref.\cite{nos2006}, so that the chance of divergence dissappears.}. In fact, it 
is proved in a more general theory, where a term $1/F^2$ is added to Eq.(\ref{action}), that
this pathology definitely dissapears.\cite{cycling-universe-paper} Aside from that, 
Eq.(\ref{actual-redshift}) makes it evident that the NLED correction is already ``built-in'' 
in the cosmic redshift estimated, for instance, from the SN host-galaxy absorption lines. 
In other words, after estimating the $B$-field strength of the host-galaxies (typically 
$10^8$-$10^9$~G) of each of the already observed SNIa, the actual redshift to be plotted 
in the Hubble diagram ($\mu(z)$~{\rm vs.}~$ z$) is going to be the effective redshift 
$(1+\tilde{z})$ discounting the correction factor $\Delta$ provided by NLED (examples in 
plots c, d in Fig.1).  


From this analysis one concludes that in case the NLED theory for the photon interaction (in a 
vacuum) with extragalactic background magnetic fields be realized in nature, it would become 
evident that any conclusion on actual cosmological redshifts drawn from SNIa or/and observations 
in the optical of GRBs afterglows had to be revised. In any case, any general NLED theory will 
lead to an effective cosmic redshfit $(1+\tilde{z}) = (1 + z)~\Delta$. In the 
specific case of $L(F) = F + 1/F$, wondering whether the theory has something to do with nature 
rests on future experiments and/or observations like observing absorption lines in  both a 
supernova and a gamma-ray burst afterglow in SN/GBRs related events, a connection that is by 
now conclusively demonstrated for a number of cases \cite{woosley-bloom2006}.


{ \textbf{Acknowledgements:}{\; HJMC thank Prof. R Ruffini and ICRANet 
Coordinating Centre, Pescara, Italy, for hospitality during preparation 
of this paper. JMS and MN acknowledge financial support from CNPq and 
FAPERJ.  SE P\'erez Bergliaffa is thanked for continued collaboration 
on this research, and G Avenda\~no Franco is thanked for computer 
assistance.  } }



\begin{thebibliography}{99}
\bibitem{gamma} Notice that on dimensional grounds,
$
\gamma = \hbar ^2\times \nu^{2},
$
where $\mu$ is a fundamental constant with dimension of length$^{-1}$.
\end{thebibliography}
\end{document}